\documentclass[aps,pra,twocolumn,superscriptaddress,bibnotes]{revtex4-2}
\usepackage{CJKutf8}
\usepackage{graphics}
\usepackage[normalem]{ulem}
\usepackage{amssymb}
\usepackage{color}
\usepackage{graphicx}
\usepackage{amsthm}
\usepackage{amsmath}
\usepackage{paralist}
\usepackage{verbatim}
\usepackage{chemarr}
\usepackage{braket}
\usepackage{enumitem}
\usepackage{float}
\usepackage{tikzit}

\tikzstyle{dotted line}=[-, style=dotted, tikzit draw=brown]
\tikzstyle{dashed line}=[-, style=dashed, tikzit draw=cyan]
\tikzstyle{green fill line}=[-, fill={green!90!black}, tikzit draw=green]
\tikzstyle{blue fill}=[-, fill=blue, tikzit fill=blue, tikzit draw={rgb,255: red,102; green,117; blue,255}]
\tikzstyle{thick red}=[-, draw=red, tikzit draw=red, line width=1pt]
\tikzstyle{thick blue}=[-, draw={rgb,255: red,28; green,176; blue,255}, tikzit draw=blue, line width=1pt]
\tikzstyle{thick black}=[-, draw=black, tikzit draw=black, line width=1pt]
\tikzstyle{dotted thick red}=[-, line width=1pt, draw=red, style=dotted, tikzit draw=red]
\tikzstyle{dotted thick blue}=[-, draw={rgb,255: red,28; green,176; blue,255}, tikzit draw=blue, line width=1pt, style=dotted]
\tikzstyle{dashed thick blue}=[-, draw={rgb,255: red,28; green,176; blue,255}, tikzit draw={rgb,255: red,83; green,19; blue,156}, line width=1pt, style=dashed]
\tikzstyle{dashed thick red}=[-, draw=red, tikzit draw={rgb,255: red,255; green,100; blue,10}, line width=1pt, style=dashed]

\usetikzlibrary{circuits.ee.IEC}
\usetikzlibrary{patterns}
\pgfdeclarelayer{nodelayer} 
\pgfdeclarelayer{edgelayer}   
\pgfsetlayers{nodelayer,edgelayer} 

\usepackage[caption=false]{subfig}
\usepackage{commath}
\usepackage{hyperref}
\newtheorem{theorem}{Theorem}

\newcommand{\spc}[1]{\mathcal{#1}}

\def\>{\rangle}
\def\<{\langle}

\newcommand{\ketbra}[2]{\ket{#1} \!  \bra{#2}}
\newcommand{\st}[1]{\mathbf{#1}}

\newcommand{\Vac}[1]{\mathrm{Vac}}

\newcommand{\map}[1]{\mathcal{#1}}

\usepackage{xr}

\begin{document}
\begin{CJK*}{UTF8}{gbsn}

\title{Quantum  networks with coherent routing of information through multiple nodes}

\author{{Hl\'er Kristj\'ansson}}
\email{Correspondence: hler.kristjansson@perimeterinstitute.ca;\\ Current address: Perimeter Institute for Theoretical Physics, 31 Caroline Street North, Waterloo, Ontario, Canada, and Institute for Quantum Computing, University of Waterloo, 200 University Avenue West, Waterloo, Ontario, N2L 3G1, Canada}
\affiliation{Quantum Group, Department of Computer Science, University of Oxford, Wolfson Building, Parks Road, Oxford, United Kingdom}
\affiliation{HKU-Oxford Joint Laboratory for Quantum Information and Computation}

\author{Yan Zhong (钟妍)}
\email{Current address: Department of Computer Science, Johns Hopkins University, 3400 North Charles Street, Baltimore, MD, United States of America}
\affiliation{QICI Quantum Information and Computation Initiative, Department of Computer Science, The University of Hong Kong, Pok Fu Lam Road, Hong Kong}

\author{Anthony Munson}
\affiliation{Department of Physics, University of Maryland, College Park, MD, United States of America}
\affiliation{Quantum Group, Department of Computer Science, University of Oxford, Wolfson Building, Parks Road, Oxford, United Kingdom}

\author{Giulio Chiribella}
\email{Correspondence: giulio.chiribella@cs.ox.ac.uk}
\affiliation{QICI Quantum Information and Computation Initiative, Department of Computer Science, The University of Hong Kong, Pok Fu Lam Road, Hong Kong}
\affiliation{Quantum Group, Department of Computer Science, University of Oxford, Wolfson Building, Parks Road, Oxford, United Kingdom}
\affiliation{HKU-Oxford Joint Laboratory for Quantum Information and Computation}
\affiliation{Perimeter Institute for Theoretical Physics, 31 Caroline Street North, Waterloo, Ontario, Canada}

\begin{abstract}

Large-scale communication networks, such as the internet, rely on routing packets of data through multiple intermediate nodes to transmit information from a sender to a receiver. In this paper, we develop a model of a quantum communication network that routes information simultaneously along multiple paths  passing through  intermediate stations.  We demonstrate that a quantum routing approach can in principle extend the distance over which information can be transmitted reliably.   
Surprisingly, the benefit of quantum routing  also applies to the transmission of classical information:  even if the transmitted data is purely classical,   delocalising it on multiple routes can enhance the achievable transmission distance. Our findings highlight the potential of a future quantum internet not only for achieving secure quantum communication and distributed quantum computing but also for extending the range of classical data transmission. 
\end{abstract}

\maketitle
\end{CJK*}

\section{Introduction}
\label{sec:introduction}

Quantum communication has the potential to revolutionise traditional telecommunication networks, enabling the transmission of private messages through quantum key distribution \cite{BennettCh1984,ekert1991quantum} and the implementation of distributed quantum computing \cite{cirac1999distributed,wehner2018quantum}.
However, noise and decoherence pose a major challenge, exponentially suppressing the transmission of quantum states over large distances. To address this challenge, quantum repeaters \cite{sangouard2011quantum} have been proposed as intermediate stations that achieve long-distance entanglement through entanglement swapping \cite{zukowski1993event} combined with entanglement purification  and error correction techniques \cite{dur1999quantum}. Nevertheless, the rate of quantum communication remains an issue, as repeated rounds of transmission are required to establish long-distance entanglement. This issue has led to a search for methods to improve the quality of long-distance communication,  either by directly reducing noise in transmission lines or by using encoding strategies to approach the optimal achievable rates \cite{shapiro2009quantum,kristjansson2021quantum}.

 A promising way to enhance quantum communication is to exploit the ability of quantum  particles to  propagate simultaneously through multiple paths,  as in the iconic double slit experiment \cite{feynman1964sands}. 
  Groundbreaking work by Gisin, Linden, Massar, and Popescu \cite{gisin2005error} has demonstrated that path superposition  could be used to filter out errors,  in principle enabling secure key distribution in highly noisy environments \cite{lamoureux2005experimental}.   More recent studies have shown enhancements in classical and quantum communication capacities \cite{abbott2018communication,chiribella2019shannon2q,kristjansson2021witnessing}. While these studies focused on direct communication between a sender and receiver, in real-world classical communication networks information is typically routed through intermediate nodes. In this setting, a crucial problem is to find the optimal route  \cite{schwartz1980routing}.
 With the emergence of quantum communication networks, such as the quantum internet \cite{kimble2008quantum}, the path of information carriers could be coherently controlled by quantum routers \cite{zhan2014perfect,caleffi2017optimal,pemberton2011perfect,zhou2013quantum,chudzicki2010parallel,yan2014single}, enabling the superposition of multiple routes. However, despite the potential benefits, the potential of quantum routing through multiple intermediate nodes remains largely unexplored.

In this paper, we introduce a model for quantum communication networks that allow for coherent information routing assisted by local operations at multiple intermediate nodes.  We demonstrate that this model  can  increase the distance over which information can be transmitted reliably in a number of  scenarios.  
In the idealised case where the path degree of freedom is not affected by decoherence, we find that the model even enables classical communication at a finite rate over arbitrarily long distances.  Our findings show that a quantum internet could be used not only for achieving distributed quantum computing and secure quantum communication,  but also for extending the range of classical communication:  even if the data is purely classical, the ability to delocalise its route can benefit the transmission.

\section{Results}
\label{sec:results}


{\bf Quantum networks with coherent control.} Here we formulate a  model  of  communication networks with quantum control over the trajectories connecting a sender to a receiver through intermediate nodes.   

\begin{figure}
\centering
\includegraphics[width=0.45\textwidth]{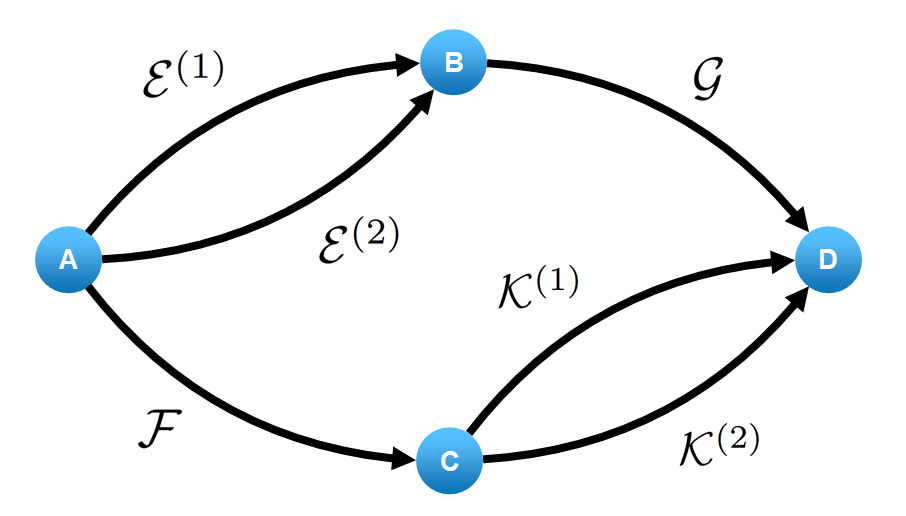}
{\caption{\label{fig:network}    Quantum  network with coherent control of the communication paths.   The vertices represent the communicating parties, while the arrows represent the available  communication channels. When multiple communication paths are available, the information carriers can propagate  in a superposition of alternative paths. 
}}
\end{figure}

In the basic scenario, illustrated in Figure \ref{fig:network},  the communication  network is described by a directed   graph, where the vertices represent communicating parties and the edges represent communication channels between them.   When a communication  channel is not used, we will model its input as being the `vacuum state', a state orthogonal to all the states used to encode information.   This modelling is inspired by quantum communication with polarisation qubits, where the information is encoded in the subspace generated by the one-photon   states $| 1\>_{   \st k,   H } \otimes | 0\>_{   \st k,   V } $ and  $| 0\>_{   \st k,   H } \otimes | 1\>_{   \st k,   V }$, 
while  the absence  of inputs is described by the vacuum state  $| 0\>_{   \st k,   H } \otimes | 0\>_{   \st k,   V } $, where $\st k$ denotes the wave vector and $H (V)$ horizontal (vertical) polarisation.

The action of a communication  channel on the extended, particle-plus-vacuum system is specified by a quantum channel (completely positive trace-preserving map)  $\widetilde{\map E}$, acting on density matrices on the Hilbert space  $\widetilde{\spc{H}} := \spc{H} \oplus \spc{H}_\textrm{Vac}$, where $\spc{H}$ is the single-particle Hilbert space, and $\spc{H}_\textrm{Vac}$ the  one-dimensional Hilbert space spanned by the vacuum  state $\ket{\textrm{vac}}$.  The channel $\widetilde{\map E}$ is an extension of the original channel ${\map E}$, acting only on the single-particle subspace.  This extension, called the vacuum extension, has Kraus operators  of the form $\{ \widetilde{E}_i := E_i +   \alpha_i  \,  |{\rm vac}\>\,\<  {\rm vac} | \}$, where  $\{  E_i\}$ are Kraus operators of the original channel $\map E$, and $\{\alpha_i\}$ are complex amplitudes satisfying the normalization condition
$\sum_{i=0}^{r-1} \, |   \alpha_i|^2   =  1$ \cite{chiribella2019shannon2q}. 

Superpositions of different network paths, for example between nodes $A$ and $B$ in Figure \ref{fig:network},  are realised by  superposing single-particle states of the form $  |\psi\>_1  \otimes  |{\rm vac}\>_2$ and $|{\rm vac}  \>_1\otimes |\psi\>_2$, 
where the subscripts 1 and 2 refer to the input ports of  the channels $\widetilde{\map E}^{(1)}$ and $\widetilde{\map E}^{(2)}$. Equivalently, the transmission can be modelled  in terms of  an external degree of freedom, which controls 
the particle's path. In this picture, the states  $  |\psi\>_1  \otimes  |{\rm vac}\>_2$ and $|{\rm vac}  \>_1\otimes |\psi\>_2$ are represented as $|\psi\> \otimes |0\>$ and $|\psi\>\otimes |1\>$, where $|0\>$ and $|1\>$ are orthogonal states of the path degree of freedom. 
The  evolution of a particle  travelling from node $A$ to node $B$ of Figure \ref{fig:network} is described by a quantum channel $\map S_{\widetilde{\map E}^{(1)},\widetilde{\map E}^{(2)}}$ with  Kraus operators \cite{chiribella2019shannon2q} 
\begin{equation}\label{eq:sup_Kraus}
{S}_{ij}   =    E^{(1)}_{i}  \, \alpha_j^{(2)}  \otimes |0\>\<0| + E_{j}^{(2)} \, \alpha_i^{(1)}  \otimes |1\>\<1| \, ,    
\end{equation}
where $E^{(k)}_{i} ,  \alpha^{(k)}_i $ are the Kraus operators and  amplitudes associated with channel ${\map E}^{(k)}$, for $k = 1,2$, and 
$|0\>$ and $|1\>$ are orthogonal states of the path degree of freedom.

If  the external degree of freedom is initially in the superposition state $\ket{+} = (\ket{0}+\ket{1})/\sqrt{2}$, the effective evolution from node $A$ to node $B$ is described by the channel $\map S_{\widetilde{\map E}^{(1)},\widetilde{\map E}^{(2)}}^{\ketbra{+}{+}}   (\rho) :  = \map S_{\widetilde{\map E}^{(1)},\widetilde{\map E}^{(2)}}  (\rho 
\otimes {\ketbra{+}{+}})$, which can be regarded as a quantum superposition of the two channels ${\mathcal{E}}^{(1)}$ and ${\mathcal{E}}^{(2)}$ \cite{Aharonov1990,oi2003interference,aaberg2004subspace,aaberg2004operations,abbott2018communication,chiribella2019shannon2q}. 
 When the two channels are identical (${\mathcal{E}}^{(1)} ={\mathcal{E}}^{(2)} \equiv \widetilde{\map E})$ 
the particle's state  after the transmission  is 
\begin{equation} \label{eq:Fsup_gen}
\begin{split}
\map S^{\ketbra{+}{+}}_{\mathcal{\widetilde{E}}}  (\rho )  
&=    \frac{ \map E \! \left( \rho \right)  + F  \rho  F^\dag   }2  \otimes \ketbra{+}{+} \\ 
&+  \frac{  \map E \! \left( \rho \right)     -  F\rho F^\dag}2 \otimes \ketbra{-}{-}     ,
\end{split}
\end{equation}  
where we use the shorthand $\map S^{\ketbra{+}{+}}_{\mathcal{\widetilde{E}}} := \map S^{\ketbra{+}{+}}_{\mathcal{\widetilde{E}},\mathcal{\widetilde{E}}}$, the first (second) factor of each tensor product
corresponds to the message (path), and $F  :=  \sum_i  \bar{\alpha}_{i} \, {E}_{i}$ is called the vacuum interference operator  \cite{chiribella2019shannon2q} (a summary of the derivation is also provided in the Methods section \ref{sec:methods}).

Given two nodes of a quantum communication network,  the problem is to find the  optimal way to transfer  information from one node  to the other.    This amounts to optimising the route through the intermediate nodes,  the encoding operations performed by the sender, the decoding operations performed by the receiver, and the operations performed  by the intermediate parties.   

With  the general framework in place, in the next sections we will show that the superposition of network paths travelling through multiple intermediate nodes can enhance the rate for the transmission of classical data.

\medskip 

{\bf Communication through asymptotically long sequences  of binary asymmetric channels.}     An important type of classical channel is the binary asymmetric channel, which transmits a bit with different probabilities of error depending on whether the bit value is $0$ or $1$. The quantum realisation of this channel is given by the map
\begin{align}
\nonumber 	\map E (\rho)   =  & (1-q)  \,  |0\>\<0|  \,  \< 0| \rho |0\>  +  q  \,  |1\>\<1| \, \< 0| \rho |0\> \\
  &  + p  \,  |0\>\<0|  \,  \< 1| \rho |1\>  +  (1-p) \,  |1\>\<1| \,  \<1|\rho|1\>  \, .  \label{eq:asymmetric}
\end{align}
 In the following, we shall consider first the special case of binary asymmetric channel in which the  value 0 is transmitted without error, namely $q=0$.  The resulting channel is known as the $Z$-channel \cite{golomb1980limiting,moskowitz1996analysis} and has several applications in optical communications \cite{baumert1979coding}. The quantum realisation of the $Z$-channel is  the map 
 \begin{align}	\map E (\rho)   =   p \, |0\>\<0| +  (1-p) \,  \rho_{\rm diag} \, ,  \label{eq:Z}
\end{align}
with $\rho_{\rm diag}  :  =  |0\>\<0|  \,  \<0| \rho |0\>  +  |1\>\<1|  \,  \<1|  \rho |1\>$.  
 
Now, suppose that a quantum particle has to traverse $n$ $Z$-channels on the way from the sender to the receiver.  If the particle follows a definite path,  its state after the transmission will be
\begin{equation}\label{eq:ZN}
	\map E^n (\rho) = \left[ 1-(1-p)^n \right] \ketbra{0}{0} + (1-p)^n \rho_{\rm diag} \, ,
\end{equation}
which is equivalent to another $Z$-channel with effective error probability $ 1-(1-p)^n$.     
For large $n$, the dependence of the output on the input vanishes exponentially, resulting in an exponential decay of the transmission rate. 

\begin{figure*}
	\centering
	\begin{tikzpicture}
	\begin{pgfonlayer}{nodelayer}
		\node [style=none] (0) at (-7.5, -1) {};
		\node [style=none] (1) at (-7.5, -2) {};
		\node [style=none] (10) at (-6.5, 0) {};
		\node [style=none] (11) at (-6.5, -1.5) {};
		\node [style=none] (12) at (-6.5, 2) {};
		\node [style=none] (13) at (-6.5, -2) {};
		\node [style=none] (14) at (-5, 2) {};
		\node [style=none] (15) at (-5, -2) {};
		\node [style=none] (16) at (-5, 1.25) {};
		\node [style=none] (17) at (-5, -1.25) {};
		\node [style=none] (18) at (-3.5, -1.25) {};
		\node [style=none] (19) at (-3.5, 1.25) {};
		\node [style=none] (20) at (-3.5, 2) {};
		\node [style=none] (21) at (-3.5, 0.5) {};
		\node [style=none] (22) at (-3.5, -0.5) {};
		\node [style=none] (23) at (-3.5, -2) {};
		\node [style=none] (24) at (-1.5, 2) {};
		\node [style=none] (25) at (-1.5, 0.5) {};
		\node [style=none] (26) at (-1.5, -0.5) {};
		\node [style=none] (27) at (-1.5, -2) {};
		\node [style=none] (28) at (-3.5, 0.5) {};
		\node [style=none] (30) at (-1.5, 1.25) {};
		\node [style=none] (31) at (-1.5, -1.25) {};
		\node [style=none] (53) at (3, -1.25) {};
		\node [style=none] (54) at (3, 1.25) {};
		\node [style=none] (55) at (3, 2) {};
		\node [style=none] (56) at (3, 0.5) {};
		\node [style=none] (57) at (3, -0.5) {};
		\node [style=none] (58) at (3, -2) {};
		\node [style=none] (59) at (5, 2) {};
		\node [style=none] (60) at (5, 0.5) {};
		\node [style=none] (61) at (5, -0.5) {};
		\node [style=none] (62) at (5, -2) {};
		\node [style=none] (63) at (3, 0.5) {};
		\node [style=none] (65) at (5, 1.25) {};
		\node [style=none] (66) at (5, -1.25) {};
		\node [style=none] (115) at (11, -1.25) {};
		\node [style=none] (116) at (11, 1.25) {};
		\node [style=none] (117) at (11, 2) {};
		\node [style=none] (118) at (11, 0.5) {};
		\node [style=none] (119) at (11, -0.5) {};
		\node [style=none] (120) at (11, -2) {};
		\node [style=none] (121) at (13, 2) {};
		\node [style=none] (122) at (13, 0.5) {};
		\node [style=none] (123) at (13, -0.5) {};
		\node [style=none] (124) at (13, -2) {};
		\node [style=none] (125) at (11, 0.5) {};
		\node [style=none] (126) at (14.5, 1.25) {};
		\node [style=none] (127) at (13, 1.25) {};
		\node [style=none] (128) at (13, -1.25) {};
		\node [style=none] (129) at (14.5, -1.25) {};
		\node [style=none] (130) at (14.5, 2) {};
		\node [style=none] (131) at (14.5, -2) {};
		\node [style=none] (132) at (16, -2) {};
		\node [style=none] (133) at (16, 2) {};
		\node [style=none] (134) at (16, 0) {};
		\node [style=none] (135) at (16, -1.5) {};
		\node [style=none] (168) at (-8, -1.5) {};
		\node [style=none] (169) at (-8, -1.5) {$+$};
		\node [style=none] (175) at (-7.5, -1.5) {};
		\node [style=none] (182) at (4, 1.25) {};
		\node [style=none] (183) at (4, -1.25) {};
		\node [style=none] (187) at (12, 1.25) {};
		\node [style=none] (188) at (12, -1.25) {};
		\node [style=none] (206) at (12, 1.25) {$\widetilde{\map{E}}_n$};
		\node [style=none] (207) at (12, -1.25) {$\widetilde{\map{E}}_n$};
		\node [style=none] (228) at (-2.5, 1.25) {$\widetilde{\map{E}}_1$};
		\node [style=none] (229) at (-2.5, -1.25) {$\widetilde{\map{E}}_1$};
		\node [style=none] (239) at (4, 1.25) {$\widetilde{\map{E}}_2$};
		\node [style=none] (240) at (4, -1.25) {$\widetilde{\map{E}}_2$};
		\node [style=none] (248) at (-8.5, 0) {};
		\node [style=none] (249) at (18, 0) {};
		\node [style=none] (250) at (18, -1.5) {};
		\node [style=none] (255) at (-7.5, 0.75) {$\spc{H}$};
		\node [style=none] (257) at (6.5, 1.25) {};
		\node [style=none] (258) at (10.25, 1.25) {};
		\node [style=none] (262) at (-7.5, -2.5) {$\spc{H}_P$};
		\node [style=none] (268) at (6.5, -1.25) {};
		\node [style=none] (269) at (10.25, -1.25) {};
		\node [style=none] (270) at (17, 0.75) {$\spc{H}$};
		\node [style=none] (271) at (17, -2.25) {$\spc{H}_P$};
		\node [style=none] (272) at (0, -1.25) {};
		\node [style=none] (273) at (0, 1.25) {};
		\node [style=none] (274) at (0, 2) {};
		\node [style=none] (275) at (0, 0.5) {};
		\node [style=none] (276) at (0, -0.5) {};
		\node [style=none] (277) at (0, -2) {};
		\node [style=none] (278) at (1.5, 2) {};
		\node [style=none] (279) at (1.5, 0.5) {};
		\node [style=none] (280) at (1.5, -0.5) {};
		\node [style=none] (281) at (1.5, -2) {};
		\node [style=none] (282) at (0, 0.5) {};
		\node [style=none] (283) at (1.5, 1.25) {};
		\node [style=none] (284) at (1.5, -1.25) {};
		\node [style=none] (288) at (0.75, -1.25) {$\widetilde{\map{R}}_1$};
		\node [style=none] (289) at (0.75, 1.25) {$\widetilde{\map{R}}_1$};
		\node [style=none] (290) at (6.5, -1.25) {};
		\node [style=none] (291) at (6.5, 1.25) {};
		\node [style=none] (292) at (6.5, 2) {};
		\node [style=none] (293) at (6.5, 0.5) {};
		\node [style=none] (294) at (6.5, -0.5) {};
		\node [style=none] (295) at (6.5, -2) {};
		\node [style=none] (296) at (8, 2) {};
		\node [style=none] (297) at (8, 0.5) {};
		\node [style=none] (298) at (8, -0.5) {};
		\node [style=none] (299) at (8, -2) {};
		\node [style=none] (300) at (6.5, 0.5) {};
		\node [style=none] (301) at (8, 1.25) {};
		\node [style=none] (302) at (8, -1.25) {};
		\node [style=none] (303) at (7.25, -1.25) {$\widetilde{\map{R}}_2$};
		\node [style=none] (304) at (7.25, 1.25) {$\widetilde{\map{R}}_2$};
		\node [style=none] (305) at (8.75, 1.25) {};
		\node [style=none] (306) at (8.75, -1.25) {};
		\node [style=none] (307) at (16, 0) {};
		\node [style=none] (308) at (14.5, -1.25) {};
	\end{pgfonlayer}
	\begin{pgfonlayer}{edgelayer}
		\draw [in=90, out=-90, looseness=3.25] (0.center) to (1.center);
		\draw [in=-180, out=180, looseness=3.75] (0.center) to (1.center);
		\draw (12.center) to (13.center);
		\draw (13.center) to (15.center);
		\draw (12.center) to (14.center);
		\draw (14.center) to (15.center);
		\draw (20.center) to (28.center);
		\draw (24.center) to (25.center);
		\draw (28.center) to (25.center);
		\draw (20.center) to (24.center);
		\draw (22.center) to (26.center);
		\draw (22.center) to (23.center);
		\draw (23.center) to (27.center);
		\draw (26.center) to (27.center);
		\draw (55.center) to (63.center);
		\draw (59.center) to (60.center);
		\draw (63.center) to (60.center);
		\draw (55.center) to (59.center);
		\draw (57.center) to (61.center);
		\draw (57.center) to (58.center);
		\draw (58.center) to (62.center);
		\draw (61.center) to (62.center);
		\draw (117.center) to (125.center);
		\draw (121.center) to (122.center);
		\draw (125.center) to (122.center);
		\draw (117.center) to (121.center);
		\draw (119.center) to (123.center);
		\draw (119.center) to (120.center);
		\draw (120.center) to (124.center);
		\draw (123.center) to (124.center);
		\draw (130.center) to (131.center);
		\draw (130.center) to (133.center);
		\draw [in=90, out=-90] (133.center) to (132.center);
		\draw (131.center) to (132.center);
		\draw [style=thick red] (16.center) to (19.center);
		\draw [style=thick red] (127.center) to (126.center);
		\draw [style=thick red] (257.center) to (65.center);
		\draw [style=thick red] (258.center) to (116.center);
		\draw [style=thick blue] (17.center) to (18.center);
		\draw [style=thick blue] (66.center) to (268.center);
		\draw [style=thick blue] (269.center) to (115.center);
		\draw [style=thick blue] (128.center) to (129.center);
		\draw [style=dashed thick red, in=180, out=0] (10.center) to (16.center);
		\draw [style=dashed thick red, in=0, out=180] (134.center) to (126.center);
		\draw [style=dashed thick blue, in=180, out=0] (10.center) to (17.center);
		\draw (274.center) to (282.center);
		\draw (278.center) to (279.center);
		\draw (282.center) to (279.center);
		\draw (274.center) to (278.center);
		\draw (276.center) to (280.center);
		\draw (276.center) to (277.center);
		\draw (277.center) to (281.center);
		\draw (280.center) to (281.center);
		\draw (292.center) to (300.center);
		\draw (296.center) to (297.center);
		\draw (300.center) to (297.center);
		\draw (292.center) to (296.center);
		\draw (294.center) to (298.center);
		\draw (294.center) to (295.center);
		\draw (295.center) to (299.center);
		\draw (298.center) to (299.center);
		\draw [style=thick red] (30.center) to (273.center);
		\draw [style=thick red] (283.center) to (54.center);
		\draw [style=thick red] (301.center) to (305.center);
		\draw [style=dotted thick red] (305.center) to (258.center);
		\draw [style=thick blue] (31.center) to (272.center);
		\draw [style=thick blue] (284.center) to (53.center);
		\draw [style=thick blue] (302.center) to (306.center);
		\draw [style=dotted thick blue] (306.center) to (269.center);
		\draw [style=dashed thick blue, in=0, out=-180] (307.center) to (308.center);
		\draw [style=thick black] (248.center) to (10.center);
		\draw [style=thick black] (175.center) to (11.center);
		\draw [style=thick black] (307.center) to (249.center);
		\draw [style=thick black] (135.center) to (250.center);
	\end{pgfonlayer}
\end{tikzpicture}
	\caption{\label{fig:rep} A superposition of two sequences of $n$ channels $(\map E_1,\dots \map ,E_n)$, specified by the vacuum extensions $(\widetilde{\map E}_1,\dots ,\widetilde{\map E}_n)$, with intermediate operations $(\widetilde{\map R}_1,\dots, \widetilde{\map R}_{n-1})$.	
  }  
\end{figure*}
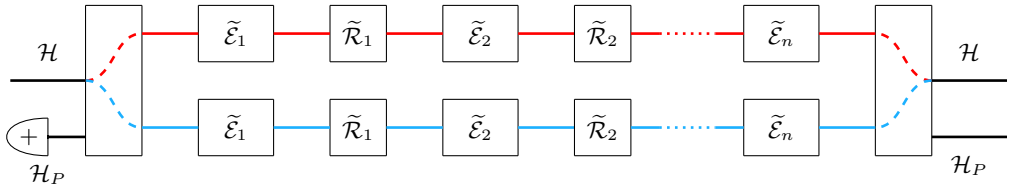

Now, suppose that two alternative paths are available.  If a quantum particle traverses them in a quantum superposition, with the path 
initialised in the $\ket{+}$ state, then
the output state is 
\begin{equation}\label{eq:Fsup_genN}
\begin{split}
\map S^{\ketbra{+}{+}}_{\mathcal{\widetilde{E}}^n}  (\rho)  
&=    \frac{ \map E^n \! \left( \rho \right)  + F^n  \rho  F^{\dag n}   }2  \otimes \ketbra{+}{+}  \\
&+  \frac{  \map E^n \! \left( \rho \right)     -  F^n \rho F^{\dag n}  }2 \otimes \ketbra{-}{-}    .
\end{split}
\end{equation}
   In the large $n$ limit,  the operator $F^n$ tends to the projector $\Pi$ on the eigenspace of $F$ associated to the eigenvalue 1 (if $F$ does not have an eigenvalue 1, then $\Pi=0$). Hence, the output state converges to   
   \begin{align}
       \nonumber     &\map S_\infty^{|+\>\<+|} (\rho)    :  = \lim_{n\to \infty}  \map S^{\ketbra{+}{+}}_{\mathcal{\widetilde{E}}^n}  (\rho)\\
      & =    \frac   {|0\>\<0|    + \Pi  \rho  \Pi  }2  \otimes \ketbra{+}{+}  + \frac    {|0\>\<0|  -  \Pi \rho \Pi}2   \otimes \ketbra{-}{-} \,.   \label{limit}
  \end{align}
 For $\Pi\not  =  0$,   this state has a non-trivial dependence on the input state $\rho$ and therefore permits a transmission of classical data, despite the fact that the paths between the sender and receiver are asymptotically long.  

The crucial question, at this point, is whether or not the operator $F$ has a non-trivial  eigenspace with  eigenvalue 1.  We now show that the answer is affirmative in 
a physical realisation of the $Z$-channel inspired by single-photon quantum optics.  Let us regard the channel as the result of two steps: first, a complete decoherence in the computational basis $\{|0\>  , |1\>\}$, and then a random operation that replaces the input with the state $|0\>$ with probability $p$, or leaves the input state invariant with probability $1-p$.  For the physical realisation of the  decoherence map,  we take a controlled-{\tt NOT}  interaction with an environment  $E_1$, initially in the state $|0\>$.  In this realisation, the vacuum extension is easily found: if the system is not present, then the environment remains unaffected. Overall, this first interaction  is described by the unitary gate
\begin{align}
\widetilde U_{\widetilde{M} E_1}=  |0\>\<0|  \otimes I_{E_1}  +  |1\>\<1|  \otimes X_{E_1}   + |{\rm vac}\>\<{\rm vac}| \otimes I_{E_1} \,,
\end{align}
where the vectors $\{|0\>, |1\>,  |\rm vac\>\}$  form a basis for the vacuum-extended message system $\widetilde{M}$, the subscript $E_1$ denotes operators acting on the environment, and  $X:=  |0\>\<1|  + |1\>\<0|$.   
Similarly, the interaction that resets  the input state to $|0\>$ can be modelled  as a {\tt SWAP} gate, where the state of the system is swapped with the state of an environment $E_2$, initially in the state $|0\>$.  The vacuum extension of the  $\tt SWAP$ gate is given by the unitary gate  
\begin{align}
 \widetilde V_{\widetilde M E_2}= \,   & {\tt SWAP}    + |{\rm vac}\>\<{\rm vac}| \otimes I_{E_2} \,.
\end{align}
Once the environments $E_1$ and $E_2$ are discarded, the above interactions give rise to an extended channel $\widetilde {\map C_1}$ with Kraus operators $\widetilde C_{10}  =  |0\>\<0|  \oplus |{\rm vac\>\<  \rm vac}|$ and $\widetilde C_{11}  =  |0\>\<1|$.  This channel is applied to the input with probability $p$. With probability $1-p$, instead, only the first interaction takes place, resulting in an extended channel $\widetilde{\map C}_2$ with Kraus operators $\widetilde C_{20}  =  |0\>\<0|  \oplus |{\rm vac\>\<  \rm vac}|$ and $\widetilde C_{21}  = |1\>\<1|$.  Overall, the evolution of the system is described by the extended channel $\widetilde {\map E}  =  p\,  \widetilde {\map C}_1  +  (1-p)\,  \widetilde {\map C}_2$. This channel has vacuum interference operator $F  =  |0\>\<0|$, and the limit channel   (\ref{limit}) becomes 
\begin{equation}\label{eq:Fsup_genN_phys}
\map S_\infty^{|+\>\<+|}  (\rho)  =  |0\>\<0  |\otimes   ( q_+ \,|+\>\<+|  +  q_-  \,  |-\>\<-|) \, ,    
\end{equation}
with $q_\pm  :  =  (1  \pm  \<0|  \rho  |0\>)/2$.  This channel
is equivalent to a classical $Z$-channel with error probability $p  = 1/2$, whose  classical capacity is known to be   $\log_2(5/4) \approx 0.32$ \cite{richardson2008modern}.  Hence, reliable classical communication through asymptotically long distances is possible.

\medskip 

{\bf Extension to variable bases.}  So far, we have considered only networks whose errors arise from identical  $Z$-channels, all defined with respect to the computational basis, and all 
with the same error probability $p$.  However, in practice, it is reasonable to expect variations within the error parameters between different parts of the network. For any chosen path through the network, the sequence of errors acting on the particle could be modelled as a sequence of possibly non-identical
$Z$-channels $\map E_n \circ \map E_{n-1} \circ \cdots \circ \map E_1$, with
\begin{equation}\label{eq:erasurek}
\map E_k (\rho) = p_k \ketbra{\eta_k}{\eta_k} + (1-p_k) \rho_{\rm diag}^{(k)} \, ,
\end{equation}
where $k \in \{1, \dots , n\}$, $p_k$ is a probability, $\{\ket{\eta_k},  \ket{\eta_{k,\perp}}\}$ is an orthonormal basis, and $\rho_{\rm diag}^{(k)}   :  =  \< \eta_k  | \rho  |\eta_k\>  \,  |\eta_k\>\<\eta_k|  +  \< \eta_{k, \perp}  | \rho  |\eta_{k, \perp}\>  \,  |\eta_{k,\perp}\>\<\eta_{k,\perp}|$\, .

For all $k$, let $\widetilde{\map E}_k$ be a vacuum extension of the channel $\map E_k$. A superposition of two identical 
sequences $\widetilde{\map E}_n \circ \widetilde{\map E}_{n-1} \circ \cdots \circ \widetilde{\map E}_1$
yields a concatenated vacuum interference operator $F=F_n F_{n-1} \cdots F_1$, where each
$F_k = \ketbra{\eta_k}{\eta_k}$
is the vacuum interference operator for $\widetilde{\map E}_k$.
Assuming that for all $k$, 
$|\braket{\eta_{k+1} | \eta_k}|<1$,
$F$ approaches zero as $n \rightarrow \infty$. One can circumvent the vacuum interference operator's decay
by inserting at each node in the network,
an intermediate operation $\widetilde{\map R}_k$
engineered to have a vacuum interference operator 
$G_k =\ketbra{\eta_{k+1}}{\eta_k}+   G_k^{\rm rest}$, with $G_k^{\rm rest} \ket{\eta_k}=0$, which is satisfied for example by the unitary $\widetilde{\map R}_k(\cdot) = \widetilde{R}_k (\cdot) \widetilde{R}_k^\dag$, with $\widetilde{R}_k = \ketbra{\eta_{k+1}}{\eta_k} + \ketbra{\eta_{k+1,\perp}}{\eta_{k,\perp}} + \ketbra{\textrm{vac}}{\textrm{vac}}$.
This construction yields
an effective vacuum interference operator for the whole sequence of $Z$-channels and intermediate operations:
$F_{\rm eff} = F_n G_{n-1} F_{n-1} \cdots G_1 F_1  $.

The superposition of two identical sequences of non-identical
 $Z$-channels concatenated with intermediate operations is
depicted in Figure \ref{fig:rep}.
If the path is initialised in the $\ket{+}$ state, then the output state of the superposed channel is
\begin{equation}\label{eq:Fsup_genN_diff}
\begin{split}
&\mathcal{S}^{\ketbra{+}{+}}_{
	\mathcal{\widetilde{E}}_{n} \mathcal{\widetilde{R}}_{n-1}\mathcal{\widetilde{E}}_{n-1} \cdots \mathcal{\widetilde{R}}_{1}\mathcal{\widetilde{E}}_{1}
}(\rho)
\\
&=    \frac{ \map E_n \! \circ \mathcal{R}_{n\!-\!1}\! \circ \map E_{n\!-\!1}\! \circ \!\cdots \! \circ  \mathcal{R}_{1} \! \circ \map E_1 \! \left( \rho \right)  + F_{\rm eff} \rho F_{\rm eff}^\dag }2  \otimes \ketbra{+}{+} 
 \\
&+    \frac{ \map E_n \! \circ \mathcal{R}_{n\!-\!1}\! \circ \map E_{n\!-\!1}\!  \circ \!\cdots \! \circ  \mathcal{R}_{1} \! \circ \map E_1 \! \left( \rho \right)  - F_{\rm eff} \rho F_{\rm eff}^\dag }2  \otimes \ketbra{-}{-}.   
\end{split}
\end{equation}
If $p_k>0$ for all $k$, 
then for large $n\rightarrow\infty$, the output state in Eq.~\eqref{eq:Fsup_genN_diff} becomes
\begin{equation}
\ketbra{\eta_n}{\eta_n} \otimes ( q_+ \,|+\>\<+|  +  q_-  \,  |-\>\<-|) \, ,
\end{equation}
where  $q_\pm  :  =  (1  \pm  \<\eta_1|  \rho  |\eta_1\>)/2$.
This expression has
the same form as 
Eq.\ \eqref{eq:Fsup_genN_phys},  enabling classical communication at a non-zero rate through asymptotically long sequences of non-identical
$Z$-channels.

\medskip

{\bf Characterisation of the channels permitting classical communication through asymptotically long sequences.}
The examples shown in the previous sections showed that it is sometimes possible to achieve a non-zero communication capacity even with infinitely long sequences of noisy channels.  We now characterise completely the set of channels giving rise to this phenomenon.

The following theorem characterises completely the set of channels that have zero capacity when used  asymptotically many times in a definite path, and yet yield finite capacity when used on a superposition of paths. (Below, $\spc{L}(\spc H)$ denotes the space of linear operators over a Hilbert space $\spc{H}$.)

\begin{theorem}
	Let  $\map E: \spc{L}(\spc H_X) \rightarrow \spc{L}(\spc H_Y)$ be a quantum channel. Let  $\widetilde{\map E}$ be a vacuum extension of $\map E$, with vacuum interference operator $F$ and Kraus representation $\{\widetilde{E}_i := \,E_i + \alpha_i \ketbra{\rm vac}{\rm vac}\}_{i=0}^{r-1}$, where $\{E_i\}_{i=0}^{r-1}$ are Kraus operators of $\map E$. Assume that for every
	intermediate operation $\map Q: \spc{L}(\spc H_Y) \rightarrow \spc{L}(\spc H_X)$, the classical capacity of the  concatenated channel  $\map E \circ (\map Q \circ \map E)^{n-1}$ tends to  zero as $n \rightarrow \infty$.
		Then, the following are equivalent:
	\begin{enumerate}
	\item There exists an intermediate operation $\map R: \spc{L}(\spc H_Y) \rightarrow \spc{L}(\spc H_X)$ with vacuum extension $\widetilde{\map R}$ and vacuum interference operator $G$, such that the superposition of two independent identical sequences of channels $\mathcal{S}^{\ketbra{+}{+}}_{
		\mathcal{\widetilde{E}} \circ  (\mathcal{\widetilde{R}} \circ \mathcal{\widetilde{E}} )^{n-1} }
		: \spc{L}(\spc H_X) \rightarrow \spc{L}(\spc H_Y \otimes \spc H_P)$
		has classical capacity strictly greater than zero as $n \rightarrow \infty$.
	\item The vacuum interference operator $F$ has singular value 1.
	\item There exist two pure states $|\phi\>\in \mathcal{H}_X$, $|\zeta\>\in \mathcal{H}_Y$ and $\theta \in [0, 2 \pi[$ such that for all $j \in \{0,\dots,r-1\}$: (a) $\alpha_j = e^{i \theta} \sqrt{\<\phi|E_j^\dag E_j|\phi\>}$ and (b) $E_j|\phi\> = \alpha_j|\zeta\>$ [and hence $\map E(|\phi\>\<\phi|) = |\zeta\>\<\zeta|$].
\end{enumerate}
\end{theorem}

The proof is given in Appendix \ref{app:proof_theorem1}.   Note that in general, $F$ is only required to have singular value 1, not necessarily an eigenvalue 1, as in the previous examples. As shown by the above theorem, the channels that permit a transmission of classical information even in  the asymptotic limit of infinitely many sequential repetitions are a very special subset of the set of all quantum channels.    In the following we will show that, nevertheless,   the superposition of paths can generically offer advantages in  non-asymptotic scenarios involving a finite number of repetitions.
\medskip


{\bf Decoherence on the path and general channels.} The possibility of  communication at a finite rate  through  asymptotically long paths is  important as a proof of principle. On the other hand, its practical applicability is limited by  two crucial assumptions: (a) that the path degree of freedom remains completely noiseless throughout the whole sequence, and (b) that the  quantum channels along the paths have  vacuum interference operators with a singular value equal to 1.  However, in practice, these assumptions are only justified as idealisations of more complex scenarios; realistic transmission lines do not in general retain perfect coherence on the path degree of freedom, nor will they always correspond to channels with unit singular values of their vacuum interference operators.

In the following, we examine what happens in the realistic case where the above assumptions are relaxed. 
We perform numerical simulations of various superpositions of binary asymmetric channels to calculate lower bounds to their classical capacity. To this purpose, we evaluate  the maximum Holevo information of the superposition of channels over all input ensembles consisting of two orthogonal states \cite{kristjansson2021witnessing}. This gives a lower bound to the Holevo capacity, which in turn gives a lower bound to the classical capacity \cite{holevo1972mathematical}. We compare this with the analytically known expression for the classical capacity of the corresponding sequence of binary asymmetric channels without a superposition of paths. For the $Z$-channel with probability $p$, the capacity is given by $\log[1+(1-p) p^{p/(1-p)}]$ \cite{richardson2008modern}, whilst the capacity of the binary asymmetric channel with  error probabilities $q$ and $p$ is given by a similar but  longer formula, which can be found in Ref.\ \cite{moser2012error}, Eq.\ (19).

Let us start by relaxing the assumption that the path is completely noiseless. In particular, we consider a dephasing error on the path. That is, between each pair of nodes, the path qubit undergoes the dephasing channel
\begin{equation}\label{eq:deph_path}
\map D (\omega) = s Z \omega Z + (1-s) \omega \, ,
\end{equation}
where $s \in ~ [0, 1/2]$ is a probability and $\omega \in \spc{L}(\spc{H}_P)$ is the initial state of the  path. 

\begin{figure}
	\includegraphics[width=\linewidth]{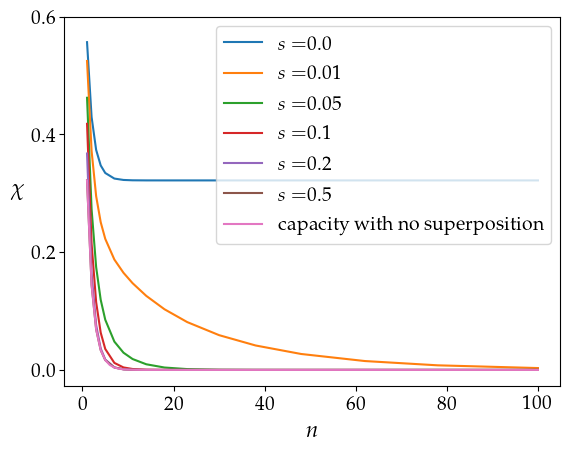}
	\caption{Lower bounds to the classical capacity ($\chi$) against sequence length ($n$) for a dephased superposition of two identical sequences of $Z$-channels, with error probability $p=0.5$, for dephasing on the path with parameter $s$, and the path degree of freedom initialised in the state $\ket{+}$. A comparison with the capacity of the $Z$-channel with error probability 0.5 without a superposition of channels is shown, highlighting that the superposition of channels gives an advantage in the capacity at small $n$ for $s \lesssim 0.1$.	\label{fig:deph_p=0.5}}
\end{figure}

We consider again two independent identical sequences of a quantum channel $\map E$, with vacuum extension $\widetilde{\map E}$ and vacuum interference operator $F$, interleaved with independent and identical dephasing channels on the path. The state of a particle after travelling through a superposition of two such sequences of channels is
\begin{equation}\label{eq:dephasing}
\begin{split}
 {\map S}^{\ketbra{+}{+}}_{\widetilde{\map E}^n ; \gamma} (\rho  ) 
&= \frac{ \mathcal{E}^n(\rho) + \gamma (F^n \rho F^{\dag n})  }{2}\otimes|+\>\<+| \\
&+\frac{\mathcal{E}^n(\rho) - \gamma (F^n \rho F^{\dag n})  }{2}\otimes|-\>\<-| \, ,
\end{split}
\end{equation}
where $\gamma = (1-2s)^n$ (see the Methods section \ref{sec:methods} for the derivation).

This shows that the magnitude of the coherence terms $\pm \gamma (F^n \rho {F^{\dagger}}^n) $ decreases exponentially with sequence length $n$. However, for finite $n$ and small enough dephasing probability $s$, communication at a non-zero rate is still possible even if the channel $\map E$ has zero capacity -- in stark contrast to using the channels in a classical configuration without a superposition of paths.

To illustrate the effect of dephasing, Figure \ref{fig:deph_p=0.5} shows a numerical plot of (lower bounds to) the classical capacity against sequence length $n$ for a dephased superposition of two identical sequences of $Z$-channels with error probability $p=0.5$, for various dephasing parameters $s$ on the path. Note that $s=0$ corresponds to no dephasing, while $s=0.5$ corresponds to complete dephasing on the path so that the final state of the path is simply the maximally mixed state (this is equivalent to using a single sequence of $n$ channels without a superposition of paths). We see that for every value of $n$, the lower bound to the capacity monotonically increases as $s$ decreases from 0.5 to 0, and an observable advantage over no superposition of paths is still present for small $n$ when $s \lesssim 0.1$.

\begin{figure*}[htb]
	\centering
	\subfloat[\label{fig:p=0.5}]{%
		\includegraphics[width=0.46\textwidth]{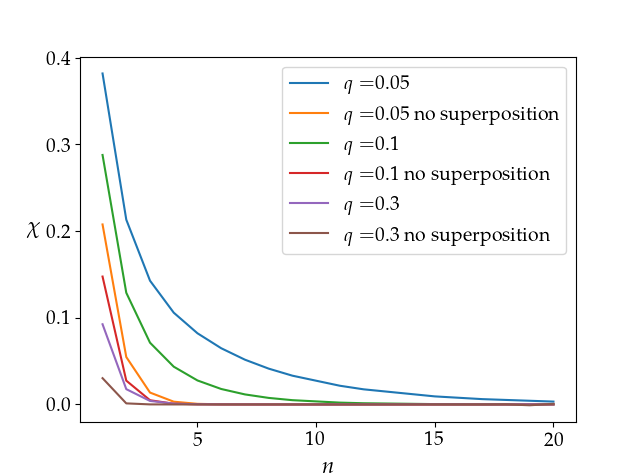}%
	} 
	\hspace{4em}
	\subfloat[\label{fig:p=0.2}]{%
		\includegraphics[width=0.46\textwidth]{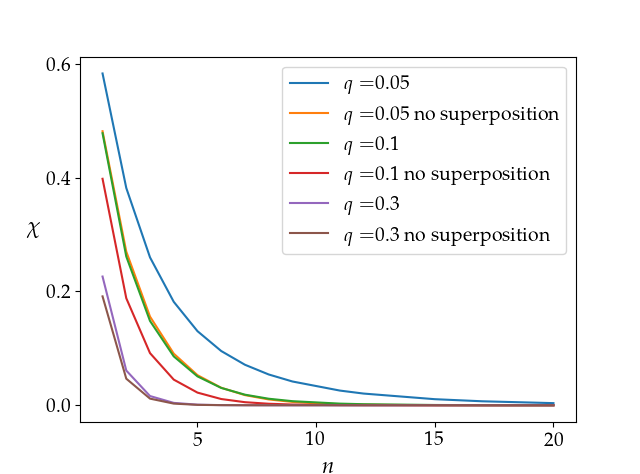}%
	}
	\caption{Lower bounds to the classical capacity ($\chi$) against sequence length ($n$) for a  superposition of two identical sequences of binary asymmetric channels, with error probabilities (a) $p=0.5$  and (b) $p=0.2$, and variable error probabilities $q$.
	The figure also shows a comparison with using the same sequence of $n$ binary asymmetric channels without a superposition of paths, where the  capacity is analytically known \cite{moser2012error}. \label{fig:decay_n}}
\end{figure*}

The second relaxation  of our initial assumptions is to consider quantum channels with non-unit singular values of their vacuum interference operator.  While in this case, it is not possible to achieve communication through asymptotically many uses of the channels, we now show that a communication advantage is still present for a finite number of uses.

To illustrate the advantage, we consider a binary asymmetric channel $\map E$ with $q  > 0$, whose vacuum extension does not give rise to unit singular values.    As in the above, we assume that the channel is implemented via a controlled-{\tt NOT} interaction with an environment initially in the state $\ket{0}$, followed by a choice of two random operations that either replace the input state with the state $\ket{0}$, with probability $p$,  or with the state $\ket{1}$, with probability $q$ (where $p+q \leq 1$). For this implementation, we obtain a vacuum extension of the binary asymmetric channel described by the Kraus operators
\begin{equation}
\begin{cases}
    \widetilde E_{1,0}  &= \sqrt{p} ( |0\>\<0|  \oplus |{\rm vac\>\<  \rm vac}|) \\
    \widetilde E_{1,1}  &= \sqrt{p} |0\>\<1| \\
    \widetilde E_{2,0}  &=  \sqrt{q} |1\>\<0| \\
    \widetilde E_{2,1}  &= \sqrt{q} |1\>\<1| \\
    \widetilde E_{2,2}  &=  \sqrt{q}  |{\rm vac\>\<  \rm vac}| \\
    \widetilde E_{3,0}  &=  \sqrt{1-p-q} ( |0\>\<0|  \oplus |{\rm vac\>\<  \rm vac}| ) \\
    \widetilde E_{3,1}  &= \sqrt{1-p-q} |1\>\<1| \, ,
\end{cases}
\end{equation}
which gives a vacuum interference operator of the form $F= (1-q)\ketbra{0}{0}$.

Figure \ref{fig:decay_n} shows (lower bounds to) the capacity as a function of $n$, for a superposition of two identical sequences of such binary asymmetric channels with error probabilities $p=0.2$ and $p=0.5$, and various values of $q$.

In order to compare the capacity to the case without a superposition of paths, we note that the concatenation of $n$ identical binary asymmetric channels $\map E^n$ with error probabilities $(q,p)$ is itself a binary asymmetric channel with error probabilities $(q_{(n)},p_{(n)})$, where
\begin{equation}
\begin{cases}
  q_{(n)} &= q \frac{1-(1-p-q)^n}{p+q}\\ 
  p_{(n)} &= p \frac{1-(1-p-q)^n}{p+q} 
  \, .
\end{cases}
\end{equation}
This enables the capacity to be calculated exactly. For each sequence of channels shown in Figure \ref{fig:decay_n}, the corresponding capacity without a superposition of paths is also shown.

For every choice of error probabilities $p$ and $q$, the coherent routing of paths provides a non-zero advantage in the classical capacity for small $n$. The gap between the capacities of the implementations with and without a superposition of paths increases as $q$ tends to 0, in the limit of which we recover the ideal case of the $Z$-channel.

\section{Discussion}

Our work provides  a new paradigm of quantum communication networks with coherent routing of information through multiple paths. 
This model  includes  earlier results that showed improvements of communication through individual channels superposed  either in space \cite{chiribella2019shannon2q,abbott2018communication} or in time \cite{kristjansson2021witnessing}. These works considered  direct communication from a sender to a receiver, possibly through multiple communication channels.  Here, instead, we address the  network scenario where information is sent through multiple intermediate nodes, undergoing a noisy channel between every two successive nodes. In this scenario, the main question is how the rate of information transmission scales with the number of nodes between the sender and the receiver.    One of our key findings is that  coherent control can sometimes be used to suppress the exponential decay of information through asymptotically long paths in the network, provided that the path is immune from decoherence.  In more realistic scenarios with small deviations from the ideal case, we still observe enhancements in the classical capacity for finite sequence lengths, compared to the case where the particle is sent on a classical trajectory.

\section{Methods}
\label{sec:methods}



{\em Superposition of quantum channels.} Here we summarise the notion of superpositions of quantum channels \cite{Aharonov1990,aaberg2004operations,aaberg2004subspace,oi2003interference,abbott2018communication,chiribella2019shannon2q} and the framework of Ref.\ \cite{chiribella2019shannon2q}.  

Consider a quantum channel (completely positive trace-preserving map) $\map E: \spc{L}(\spc{H}) \rightarrow \spc{L}(\spc{H})$, representing one use of a communication device, which can transmit a message encoded in a quantum particle. (Here $\spc{L}(\spc H)$ denotes the space of linear operators over a Hilbert space $\spc{H}$.) When a quantum particle is not transmitted, the device is described by an identity channel acting on the vacuum state. Overall, we can describe the action of the device by a \textit{vacuum extension} of the original channel:
a channel $ \widetilde{\map E} : \spc{L}(\widetilde{\spc{H}}) \rightarrow \spc{L}(\widetilde{\spc{H}}) $ is a \textit{vacuum extension} of the channel $\map E: \spc{L}(\spc{H}) \rightarrow \spc{L}(\spc{H})$ if it satisfies
\begin{equation}
\begin{cases}
\widetilde{\map E}(\rho) = \map E (\rho) ~~~~\forall \rho \in \spc{L}(\spc{H})\\
\widetilde{\map E
}(\ketbra{\textrm{vac}}{\textrm{vac}}) =\ketbra{\textrm{vac}}{\textrm{vac}} \, ,
\end{cases}
\end{equation}
where $\spc{H}$ is the single-particle Hilbert space, $\spc{H}_\textrm{Vac}$ is the  one-dimensional Hilbert space spanned by the vacuum  state $\ket{\textrm{vac}}$, and $\widetilde{\spc{H}} := \spc{H} \oplus \spc{H}_\textrm{Vac}$.
This gives the Kraus decomposition of a vacuum extension  as 
\begin{equation}\label{eq:vacext_Kraus}
\{ \widetilde{E}_i := E_i \oplus   \alpha_i  \,  |{\rm vac}\>\,\<  {\rm vac} | \}_{i=0}^{r-1} \, ,    
\end{equation}
where  $\{  E_i\}_{i=0}^{r-1}$ are Kraus operators of the original channel, and $\{\alpha_i\}_{i=0}^{r-1}$ are complex \textit{vacuum amplitudes}, satisfying
$\sum_{i=0}^{r-1} \, |   \alpha_i|^2   =  1$ \cite{chiribella2019shannon2q}.

Consider now the scenario  where we want to coherently control whether a particle is transmitted through a channel ${\mathcal{E}}^{(1)}$ or a channel ${\mathcal{E}}^{(2)}$, depending on the state of a control system $P$, `the path'.
Given two channels ${\mathcal{E}}^{(1)},{\mathcal{E}}^{(2)}$, with vacuum extensions  $\widetilde{\mathcal{E}}^{(1)}, \widetilde{\mathcal{E}}^{(2)}$,
we define a \textit{superposition of the two channels, specified by the vacuum extensions} $\widetilde{\mathcal{E}}^{(1)}, \widetilde{\mathcal{E}}^{(2)}$, with the path fixed in the state  $\omega$, as the channel
\begin{equation}\label{eq:sup}
	  \mathcal {S}^{\omega}_{ \widetilde{\mathcal{E}}^{(1)}, \widetilde{\mathcal{E}}^{(2)} }  (\rho )  :=   \mathcal{U}^\dag \circ \widetilde{\mathcal{E}}^{(1)} \otimes \widetilde{\mathcal{E}}^{(2)}  \circ \mathcal{U} \, (\rho \otimes \omega) \,.
\end{equation} 
Here,  $\mathcal{U} (\cdot) := U (\cdot) U^\dagger$ is the isomorphism
$\spc{H} \otimes \spc{H}_P \to (\spc{H}^{\rm (1)} \otimes \spc{H}_\textrm{Vac}^{\rm (2)}) \oplus (  \spc{H}^{\rm (1)}_\textrm{Vac} \otimes \spc{H}^{\rm (2)})$
that identifies the message $M$ and path $P$ in the `particle picture' with the one-particle sector of the composite system $ (\spc{H}^{\rm (1)} \oplus \spc{H}_\textrm{Vac}^{\rm (1)}) \otimes (  \spc{H}^{\rm(2)}  \oplus \spc{H}^{\rm (2)}_\textrm{Vac})$  in the `mode picture'. Explicitly, $\mathcal{U}$ is
defined  by \cite{chiribella2019shannon2q}
\begin{equation}\label{eq:U}
	 \begin{split}
	        U   (  |\psi\>_M  \otimes |0\>_P)    &:=  |\psi\>_{M^{(1)} \oplus \rm Vac}   \otimes |{\rm vac}\>_{M^{(2)} \oplus \rm Vac}  \\
	    U   (  |\psi\>_M  \otimes |1\>_P)    &:=  |{\rm vac}\>_{M^{(1)} \oplus \rm Vac}  \otimes  |\psi\>_{M^{(2)} \oplus \rm Vac}   \, .
	 \end{split}
\end{equation}
Substituting Eq.\ \eqref{eq:vacext_Kraus} into Eq.\ \eqref{eq:sup} gives the  Kraus decomposition of Eq.\ \eqref{eq:sup_Kraus} for the superposition channel $ \mathcal {S}^{\omega}_{ \widetilde{\mathcal{E}}^{(1)}, \widetilde{\mathcal{E}}^{(2)} }$.



{\em Dephasing on the superposition of channels.}    Here we extend the framework of the superposition of channels to more realistic scenarios  where the path degree of freedom is subject to dephasing errors \eqref{eq:deph_path}.  As an application, we then derive  Eq.\ \eqref{eq:dephasing}.  

Mathematically, the presence of dephasing on the path degree of freedom is equivalent to dephasing between the one-particle and vacuum sectors on one branch of the superposition:
\begin{equation}
    \map D (\rho) = s (I \oplus - \! \ketbra{\rm vac}{\rm vac}) \rho (I \oplus - \! \ketbra{\rm vac}{\rm vac})  + (1-s)\rho \, ,
\end{equation} 
where $s \in [0,1/2]$. 
Then, the dephased superposition of channels is given by
\begin{equation}\label{eq:sup_deph}
	  \mathcal {S}^{\omega}_{ \widetilde{\mathcal{E}}^{(1)}, \widetilde{\mathcal{E}}^{(2)} ;(1-2s) }  (\rho )  :=   \mathcal{U}^\dag \circ \widetilde{\mathcal{E}}^{(1)} \otimes ( \map D_s \circ \widetilde{\mathcal{E}}^{(2)})  \circ \mathcal{U} \, (\rho \otimes \omega) \, .
\end{equation} 
Note that $ \map D_s $ commutes with any vacuum extended channel $\widetilde{\mathcal{E}}^{(2)}$, because any vacuum extended channel is block diagonal with respect to the one-particle/vacuum sector partition.
	
Now consider the case where the message travels in a superposition of two identical channels $\map E$, with vacuum extensions $\widetilde{\map E}$, and the path qubit is initialised in the $\ket{+}$ state. A direct calculation (e.g.\ by substituting the Kraus operators $\{E_i \oplus \alpha_i \ketbra{\rm vac}{\rm vac}\}_{i=0}^{r-1}$ of $\widetilde{\map E}$ into Eq.\ \eqref{eq:sup_deph}), reveals that
\begin{equation}
\begin{split}
     \mathcal {S}^{\ketbra{+}{+}}_{ \widetilde{\mathcal{E}};(1-2s) }  (\rho )  
    &= \frac{ \mathcal{E}(\rho) + (1-2s) (F \rho {F^{\dagger}})  }{2}\otimes|+\>\<+| \\
&+\frac{\mathcal{E}(\rho) - (1-2s) (F \rho {F^{\dagger}})  }{2}\otimes|-\>\<-| \, ,
\end{split}
\end{equation}

Applying Eq.\ \eqref{eq:sup_deph} $n$ times, we obtain the result in Eq.\ \eqref{eq:dephasing}, where we use the shorthand  $\mathcal {S}^{\ketbra{+}{+}}_{ \widetilde{\mathcal{E}} ;(1-2s) } :=  \mathcal {S}^{\ketbra{+}{+}}_{ \widetilde{\mathcal{E}}, \widetilde{\mathcal{E}};(1-2s) } $.


%

\newpage

\section*{Acknowledgements}
We acknowledge discussions with  Alastair Abbott, Mio Murao, Sergii Strelchuk, Guillaume Boisseau, Santiago Sempere Llagostera, Robert Gardner, and Kwok Ho Wan.   This work is supported by the UK Engineering and Physical Sciences Research Council (EPSRC) through grant EP/R513295/1. Research at the Perimeter Institute for Theoretical Physics is supported by the Government of Canada through the Department of Innovation, Science and Economic Development Canada and by the Province of Ontario through the Ministry of Research, Innovation and Science. This publication was made possible through the support of the grant 61466 `The Quantum Information Structure of Spacetime' (QISS) (qiss.fr) from the John Templeton Foundation. The opinions expressed in this publication are those of the authors and do not necessarily reflect the views of the John Templeton Foundation.

\section*{Author contributions}
All authors made substantial contributions to the research and to writing up the paper, and were involved in approving the final version of the paper.

\section*{Competing interests} 
The authors declare that they have no competing financial interests.

\clearpage

\section*{Appendices}

\appendix

\section{Proof of Theorem 1}\label{app:proof_theorem1}

Recall that the operator norm of an operator $A$, $\norm{A}_\infty$, equals the largest singular value of $A$.
For any vacuum interference operator $H$, $\norm{H}_\infty \leq 1$ \cite{kristjansson2019resources}.
For any intermediate operation $\map Q$ whose vacuum extension $\widetilde{\map Q}$ has a vacuum interference operator $H$, Eq.~\eqref{eq:Fsup_genN_diff} yields the superposition of two sequences $\map E \circ (\map Q \circ \map E)^{n-1}$:
\begin{equation}\label{eq:Fsup_genN_repeaters}
	\begin{split}
		\map S^{|+\>\<+|}_{ \widetilde{\map E} \circ (\widetilde{\map Q} \circ \widetilde{\map E})^{n-1} } \! \left( \rho \right)
		&= \left[ \map E \circ (\map Q \circ \map E )^{n-1} \right]  \! \! \left( \rho \right) \otimes \frac{I}{2} \\
		&\quad + F(HF)^{n-1} \rho \! \left[ F(HF)^{n-1} \right]^\dag  \otimes \frac{X}{2} \, .
	\end{split}
\end{equation}

$1 \Rightarrow 2$:
We prove the contrapositive.
Suppose $F$ does not have the singular value 1.
Then $\norm{F}_\infty < 1$, since $\norm{F}_\infty \leq 1$.
Let $\map R$ be any intermediate operation whose vacuum extension $\widetilde{\map R}$ has a vacuum interference operator $G$.
Since $\norm{G}_\infty \leq 1$ and the operator norm is sub-multiplicative, $\norm{GF}_\infty \leq \norm{G}_\infty \norm{F}_\infty < 1$.
Thus, as $n \rightarrow \infty$, $(GF)^{n-1} \rightarrow 0$ and hence $F(GF)^{n-1} \rightarrow 0$.
By Eq.~\eqref{eq:Fsup_genN_repeaters}, the superposition $\map S^{|+\>\<+|}_{ \widetilde{\map E} \circ (\widetilde{\map R} \circ \widetilde{\map E})^{n-1} }$ becomes completely incoherent in the large-$n$ limit: $\map S^{|+\>\<+|}_{ \widetilde{\map E} \circ (\widetilde{\map R} \circ \widetilde{\map E})^{n-1} } (\rho) \rightarrow  
[\map E \circ (\map R \circ \map E )^{n-1}] (\rho) \otimes (I/2)$.
By assumption, $\map E \circ (\map R \circ \map E )^{n-1}$---and hence the superposition---has a classical capacity that tends to zero as $n \rightarrow \infty$.

$2 \Rightarrow 3$:
Let us denote the Euclidean norm of any vector $|\xi\> \in \spc{H}_M$ by $\norm{|\xi\>} := \<\xi|\xi\>^{1/2}$.
Suppose $F$ has the singular value 1.
Then there exists a pure state $|\phi\>$ such that $\norm{F|\phi\>} = 1$.
Without loss of generality, assume that $E_j|\phi\> \neq 0$ and $\alpha_j \neq 0$ if $j \leq r$, $E_j|\phi\> \neq 0$ and $\alpha_j = 0$ if $r < j \leq s$, $E_j|\phi\> = 0$ and $\alpha_j \neq 0$ if $s < j \leq t$, and $E_j|\phi\> = 0$ and $\alpha_j = 0$ if $j > t$.
$\norm{F|\phi\>}$ satisfies
\begin{align}\label{eq:Fleq1}
	\norm{F|\phi\>} &= \norm{ \sum_j \bar{\alpha}_j \, E_j|\phi\> } 
	= \norm{ \sum_{j \leq r} \bar{\alpha}_j \, E_j|\phi\> } \\
	&\leq \sum_{j \leq r} \norm{ \bar{\alpha}_j \, E_j|\phi\> }
	= \sum_{j \leq r} \abs{ \bar{\alpha}_j \, \norm{ E_j|\phi\> } } \\
	&\leq \left( \sum_{j \leq r} \norm{ E_j|\phi\> }^2 \right)^{1/2} \left( \sum_{j \leq r} |\alpha_j|^2 \right)^{1/2} \\
	&\leq \left( \sum_{j \leq s} \norm{ E_j|\phi\> }^2 \right)^{1/2} \left( \sum_{j \leq t} |\alpha_j|^2 \right)^{1/2} \\
	&= 1 \, .
\end{align}
The first inequality is the triangle inequality; the second inequality is the Cauchy-Schwarz inequality; and the last equality follows because $\sum_{j \leq s} \norm{E_j|\phi\>}^2 = \sum_j \norm{E_j|\phi\>}^2 = \<\phi| \sum_j E_j^\dag E_j |\phi\> = \<\phi|\phi\> = 1$ ($\sum_i E_j^\dag E_j = I_M$) and $\sum_j |\alpha_j|^2 = 1$.

All three of the above inequalities are equalities, since $\norm{F|\phi\>} = 1$.
The third inequality's saturation implies that $r = s = t$.
The second inequality's saturation implies the existence of a $\theta \in [0,2\pi[$ such that $\alpha_j = e^{i \theta} \norm{E_j|\phi\>} = e^{i \theta} \<\phi|E_j^\dag E_j|\phi\>^{1/2}$ for all $j \leq r$; this relation holds for all $j$, since $E_j|\phi\> = 0$ and $\alpha_j = 0$ for $j > r$.
The first inequality's saturation implies that all $E_j|\phi\>$ are equal up to normalisation for $j \leq r$.
Hence, there exists a pure state $|\zeta\>$ such that $E_j|\phi\> = e^{i\theta} \norm{E_j|\phi\>} |\zeta\> = \alpha_j |\zeta\>$ for all $j \leq r$; this relation holds for all $j$, since $E_j|\phi\> = 0$ and $\alpha_j = 0$ for $j > r$.
Lastly, $\map E (|\phi\>\<\phi|) = \sum_j E_j|\phi\>\<\phi|E_j^\dag = |\zeta\>\<\zeta| \sum_j |\alpha_j|^2 = |\zeta\>\<\zeta|$.

$3 \Rightarrow 1$:
Let $|\phi\>$ be a pure state and let $\theta \in [0,2\pi[$ such that $\alpha_j = e^{i \theta} \<\phi|E_j^\dag E_j|\phi\>^{1/2}$ for all $j$, and let $|\zeta\>$ be a pure state such that $E_j|\phi\> = \alpha_j|\zeta\>$ for all $j$.
Then $F|\phi\> = \sum_j \bar{\alpha}_j E_j|\phi\> = |\zeta\> \sum_j |\alpha_j|^2 = |\zeta\>$.
Let $\map R$ be a channel induced by a unitary $U \in \spc{L}(\spc{H}_M)$ satisfying $U|\zeta\> = |\phi\>$.
The vacuum extension $\widetilde{\map R}$ induced by $U \oplus |{\rm vac}\>\<{\rm vac}|$ has the vacuum interference operator $G = U$.
For each $n$, the sequence $\widetilde{\map E} \circ (\widetilde{\map R} \circ \widetilde{\map E})^{n-1}$ has the vacuum interference operator $F(GF)^{n-1}$.
Since $GF|\phi\> = |\phi\>$, $F(GF)^{n-1}|\phi\> = |\zeta\>$.
Furthermore, there exists a state $|\phi^\perp\>$ orthogonal to $|\phi\>$ such that $F(GF)^{n-1}|\phi^\perp\> \neq \ket{\zeta}$; otherwise, $F(GF)^{n-1}[(|\phi\>+|\phi^\perp\>)/\sqrt{2}] = \sqrt{2} |\zeta\>$ which implies the impossibility $\norm{F(GF)^{n-1}}_\infty \geq \sqrt{2}$.
By assumption, $\mathcal{{E}} \circ (\mathcal{{R}} \circ \mathcal{{E}} )^{n-1}$ has a classical capacity that tends to zero as $n \rightarrow \infty$, so every output of the channel approaches some fixed state $\sigma$.
By Eq.~\eqref{eq:Fsup_genN_repeaters}, the superposition $\map S^{|+\>\<+|}_{ \widetilde{\map E} \circ (\widetilde{\map R} \circ \widetilde{\map E})^{n-1} }$ evolves a state $\rho$ to 
\begin{equation}
	\sigma \otimes \frac{I}{2} + F(GF)^{n-1}\rho \! \left[ F(GF)^{n-1} \right]^\dag \otimes \frac{X}{2} \, .
\end{equation}
In the large-$n$ limit, the superposition evolves $|\phi\>\<\phi|$, but not $|\phi^\perp\>\<\phi^\perp|$, to $|\zeta\>\<\zeta|$, yielding distinct outputs on distinct inputs.
Thus, the superposition has a classical capacity that tends to a nonzero value as $n \rightarrow \infty$.
\qed

\end{document}